\documentclass[aip, amsmath,amssymb,reprint]{revtex4-1}
\usepackage{graphicx,color,upgreek}
\usepackage{mathtools}
\usepackage{amsmath}
\usepackage{layouts}
\usepackage{float}
\usepackage{hyperref}
\hypersetup{
    colorlinks=true,
    linkcolor=blue,
    filecolor=magenta,      
    urlcolor=cyan,
    pdftitle={Overleaf Example},
    pdfpagemode=FullScreen,
    }
\usepackage{makecell}
\usepackage[abs]{overpic}
\usepackage{helvet}
\usepackage{graphicx,color,natbib}
\usepackage{placeins}
\usepackage[dvipsnames]{xcolor}
\usepackage{array, booktabs, makecell}
\usepackage{siunitx}
\usepackage{longtable}
\usepackage{tabularx}
\usepackage{booktabs}
\let\oldAA\AA
\renewcommand{\AA}{\text{\normalfont\oldAA}}
\usepackage{textcomp}
\usepackage{color}
\usepackage{amssymb,amsmath} 
\usepackage{siunitx}

\begin{document}

\title{Identification and classification of clusters of dipolar colloids in an external field}
\author{Katherine Skipper}

\affiliation{H.H. Wills Physics Laboratory, Tyndall Ave., Bristol, BS8 1TL, UK}
\affiliation{Centre for Nanoscience and Quantum Information, Tyndall Avenue, Bristol BS8 1FD, UK}

\author{Fergus J. Moore}
\affiliation{H.H. Wills Physics Laboratory, Tyndall Ave., Bristol, BS8 1TL, UK}
\affiliation{Centre for Nanoscience and Quantum Information, Tyndall Avenue, Bristol BS8 1FD, UK}

\author{C. Patrick Royall}
\affiliation{Gulliver UMR CNRS 7083, ESPCI Paris, Universit\' e PSL, 75005 Paris, France.}
\affiliation{School of Chemistry, Cantock's Close, University of Bristol, BS8 1TS, UK}
\affiliation{Centre for Nanoscience and Quantum Information, Tyndall Avenue, Bristol BS8 1FD, UK}
\affiliation{H.H. Wills Physics Laboratory, Tyndall Ave., Bristol, BS8 1TL, UK}
\email{paddy.royall@espci.psl.eu}

\begin{abstract}
Colloids can acquire a dipolar interaction in the presence of an external electric field. At high field strength, the particles form strings in the field direction. However at weaker field strength, competition with isotropic interactions is expected. One means to investigate this interplay between dipolar and isotropic interactions is to consider clusters of such particles. We have therefore identified, using the GMIN basinhopping tool, a rich library of lowest energy clusters of a dipolar colloidal system where the dipole orientation is fixed with respect to the $z$--axis, and the dipole strength is varied for $m$--membered clusters of $7 \le m \le13$. In the regime where the isotropic and dipolar interactions are comparable,  we find elongated polytetrahedral, octahedral and spiral clusters as well as a set of non--rigid clusters which emerge close to the transition to strings. We further implement a search algorithm which identifies these minimum energy clusters in bulk systems using the \emph{topological cluster classification} [\emph{J. Chem. Phys.} \textbf{139} 234506 (2013)]. We demonstrate this methodology with computer simulations which shows instances of these clusters as a function of dipole strength.
\end{abstract}

\maketitle

\section{Introduction}
\label{sectionIntroduction}

The microscopic structure of amorphous systems has long been interpreted in terms of pairwise correlations, from the work of Ornstein and Zernike a century ago onwards~\cite{hansen}. However in the case of many phenomena, such as crystal nucleation~\cite{royall2023,kawasaki2010,russo2012} and the glass transition~\cite{frank1952,tarjus2005,charbonneau2012,royall2015physrep}, higher--order correlations in structure have
been shown to be important. One way to tackle higher--order structure has its roots in the work of Sir Charles Frank who showed that the minimum energy cluster of 13 Lennard--Jones particles is an icosahedron. Frank inferred ``that this will be a very common grouping in liquids''~\cite{frank1952}.

It has since become possible to identify such icosahedra, and other geometric motifs in bulk systems~\cite{coslovich2007,malins2013tcc,tanaka2019}. By identifying minimum energy clusters for certain model systems~\cite{doye1995,wales1997} and determining bonds in a bulk system, one may identify groups of particles whose bond topology is identical to that of isolated minimum energy clusters. Therefore, in addition to isolated clusters~\cite{malins2009,meng2010,klix2013,manoharan2015} it is possible to identify geometric motifs which might be expected to be significant in the bulk~\cite{malins2013tcc,taffs2010jcp,taffs2010jpcm,malins2013isomorph}. This has been applied to crystal nucleation~\cite{taffs2016,taffs2013,gispen2023,russo2012,fiorucci2020,kawasaki2010}, the glass transition~\cite{vanblaaderen1995,campo2020,royall2017,crowther2015} and network---forming systems such as colloidal gels~\cite{royall2021,richard2018}.

So far, this sort of analysis has largely been restricted to particles with isotropic interactions. Of course very many systems have particles which interact in an anisotropic manner. Among the most simple ways to introduce anisotropy is through a dipolar interaction. For example the Stockmayer model combines a dipolar interaction with a Lennard--Jones potential~\cite{miller2005}, and in this way is a simple model for a molecule with a permanent dipole and magnetic nanoparticles~\cite{boles2016}. In colloidal systems, it is more usual that dipolar interactions are induced by an external electric or magnetic field~\cite{ivlev}. The phase behaviour of these systems has been studied in experiment~\cite{ivlev,yethiraj2003,colla2018,semwal2022} and simulation~\cite{hynninen2005}. At sufficient field strength, the particles organize into vertically--oriented ``strings''~\cite{yethiraj2003}. In addition to their fundamental interest, these \emph{electro--rheological fluids} have potential applications as smart shock absorbers and clutches as well as in photonic materials~\cite{vanblaaderen2004}.

Here we consider the minimum energy clusters of a basic model for dipolar colloids. Motivated by the comparable similarity of the higher--order structure of colloidal fluids (represented as hard spheres or particles with an effective attraction~\cite{semwal2022}) and Lennard--Jones liquids, in that the same clusters are found in both~\cite{taffs2010jcp,taffs2010jpcm,robinson2019,wales1997}, we consider a combination of a Lennard--Jones and dipolar interaction aligned along the $z$--axis. That is to say, we consider the Stockmayer model modified such that the dipolar contribution is aligned in $z$. Now the minimum energy state of two dipolar particles is alignment with the applied field. If we consider the case where more particles are added to the system, then at low dipole strengths however, the energy of the isotropic interaction must at some point be comparable to the dipolar interaction. In a study of Stockmayer particles~\citet{miller2005} discovered a rich family of clusters, with complex knots and rings occurring at low dipole strength.

We have identified the minimum energy clusters of $7 \le m \le 13$ particles for the case applicable to induced dipoles in an external field. This was achieved with the basin hopping algorithm implemented in the GMIN energy minimization package~\cite{wales1997,GMIN}. In all cases we observe a string forming at high dipole strengths. The pathway to string assembly as the dipole strength is increased reveals a set of distinct clusters. We have characterized these structures as Lennard Jones polytetrahedra, non-Lennard-Jones polytetrahedra (which tend to be stretched in the field direction), clusters based on octahedra and Bernal spirals. We also find that as the field strength is increased further, non--rigid particles are found at high fields, and finally strings are formed.

Each minimum energy cluster has a unique bond topology that can be identified by a Voronoi decomposition. To characterize particle--resolved experimental and simulation data, we have implemented these new clusters into the \emph{topological cluster classification}~\cite{malins2013tcc}. This is a computational tool which identifies target clusters by their bond topology. We finally demonstrate the validity of this approach with computer simulation results.

\begin{figure}
\includegraphics[width=35 mm]{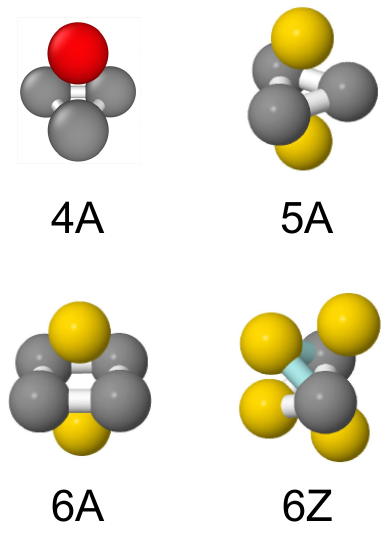}
\caption{4A, 5A, 6A and 6Z clusters. Grey indicates ring particles, yellow indicates spindle particles. Red is a sinlge spindle particle.}
\label{figBasic}
\end{figure}

This work is organized as follows. In the methodology section (Sec. ~\ref{sectionMethods}), we first outline the energy minimization methods (Sec.~\ref{sectionEnergy}) before describing briefly the topological cluster classifcation  (Sec.~\ref{sectionTopological}) and finally the molecular dynamics simulations with which we test our implementation in Sec.~\ref{sectionMolecular}. The results are presented in Sec.~\ref{sectionResults}. This is broken up into the results from the energy minimization (Sec.~\ref{sectionMinimum}) followed by the implementation of the topological cluster classification (Sec.~\ref{sectionIdentification}). This itself is divided to identification of stretched polytetrahedra (Sec.~\ref{sectionNonLennard}), clusters based on the 6A octahedron (Sec.~\ref{sectionOctahedral}) and spiral clusters (Sec.~\ref{sectionSpiral}). We then consider the results from the computer simulations in Sec.~\ref{sectionResultsMolecular}. Finally we discuss our findings in Sec.~\ref{sectionDiscussion} and conclude in Sec.~\ref{sectionConclusion}.

\section{Methods}
\label{sectionMethods}

\subsection{Energy minimisation simulation}
\label{sectionEnergy}

The system considered is a set of $m$ particles interacting through the Stockmayer potential where the dipoles are fixed parallel to the $z$--axis. The interparticle interaction is defined thus:

\begin{equation}
\beta u_\mathrm{ljdip}(r,\theta)=4\beta \epsilon\left[\left(\frac{\sigma}{r}\right)^{12}-\left(\frac{\sigma}{r}\right)^{6}\right]-\beta \mu^2 \frac{\sigma^3}{r^3}(1-3\cos^2{\theta})
\label{eqULJdip}
\end{equation}

\noindent
where $\beta=1/k_BT$, $\theta$ is the azimuthal angle between the two reduced dipole moments $\mu$. Here we set the thermal energy to unity. The minimum energy cluster of each combination of $m$ and $\mu$ was explored using the GMIN energy minimization package. A modified version of the in-built Stockmayer potential was used. GMIN uses a ``basin--hopping'' algorithm to minimize the energy~\cite{wales1997}. The basin--hopping algorithm randomly peturbs the coordinates, then performs an optimization which is rejected or accepted based on Monte--Carlo criteria. This is repeated until a specified convergence criterion is met.

Reduced units were used ($\epsilon$ and $\sigma$ are fixed to unity) and the dipole strength $\mu$ was varied between 0 and 3. 25000 basin hopping steps were performed and temperature, which determines the Monte Carlo basin hopping threshold was held at 1.5. It should be noted that as in all energy minimization simulation there can be no absolute assurance that these are the global minima, but for each value of $\mu$ and $m$, the energy minimization was repeated ten times. Only the lowest energy clusters were considered. In the case of larger clusters ($m=12,13$), for high field strengths, not all runs at the intersecting string --  string transition converged to the same shape. This is presumably due to the large interaction strengths. However, our focus in this work is on the clusters rather than the intersecting strings or strings, so we neglect this here.

\subsection{Topological cluster classification}
\label{sectionTopological}

For a complete explanation of the topological cluster classification~\citet{malins2013tcc} should be referred to. The topological cluster classification identifies target clusters by their bond network, these are polyhedra which are associated with a unique bond topology. The bond network may be defined either with a distance criterion or with a Voronoi decomposition combined with a distance criterion. As the basic building block for clusters, the algorithm constructs all the three, four and five-membered rings which can be constructed along the bond network. Rigid structures are built by considering all the particles that are bonded to every member of each ring. From these small rigid clusters, larger and more complex shapes are built. Figure \ref{figBasic} shows the three basic clusters of interest here: the 4A (tetrahedron), 6A (octahedron) and 6Z (polytetrahedron)~\footnote{Note that these differ from the basic clusters used in the previous version of the TCC~\cite{malins2013tcc}}. The grey particles indicate rings and the yellow particles are spindles. Where a particle is both a ring and a spindle they are colored for clarity. Bonds indicate rings. Red particles are bonded to three particles as part of a single--spindle 4A cluster (Fig.~\ref{figBasic}).

\begin{figure*}
\centering
\includegraphics[width=140 mm]{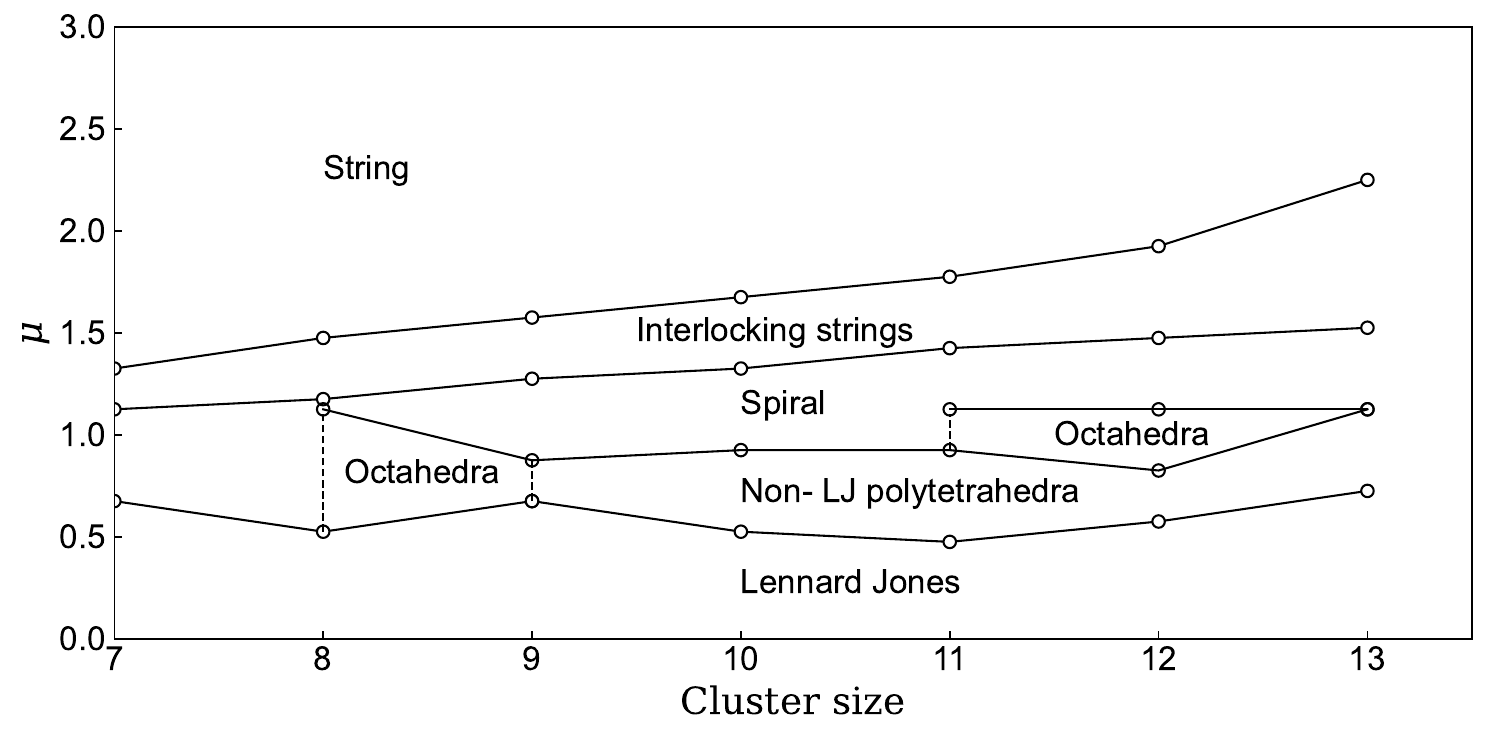}
\caption{Summary ``phase diagram'' of Lennard--Jones particles with a dipole whose direction is fixed in the $z$--direction determined using GMIN. Dotted lines indicate the change of structure.}
\label{figGMINPhase}
\end{figure*}

\begin{figure*}
\centering
\includegraphics[width=180 mm]{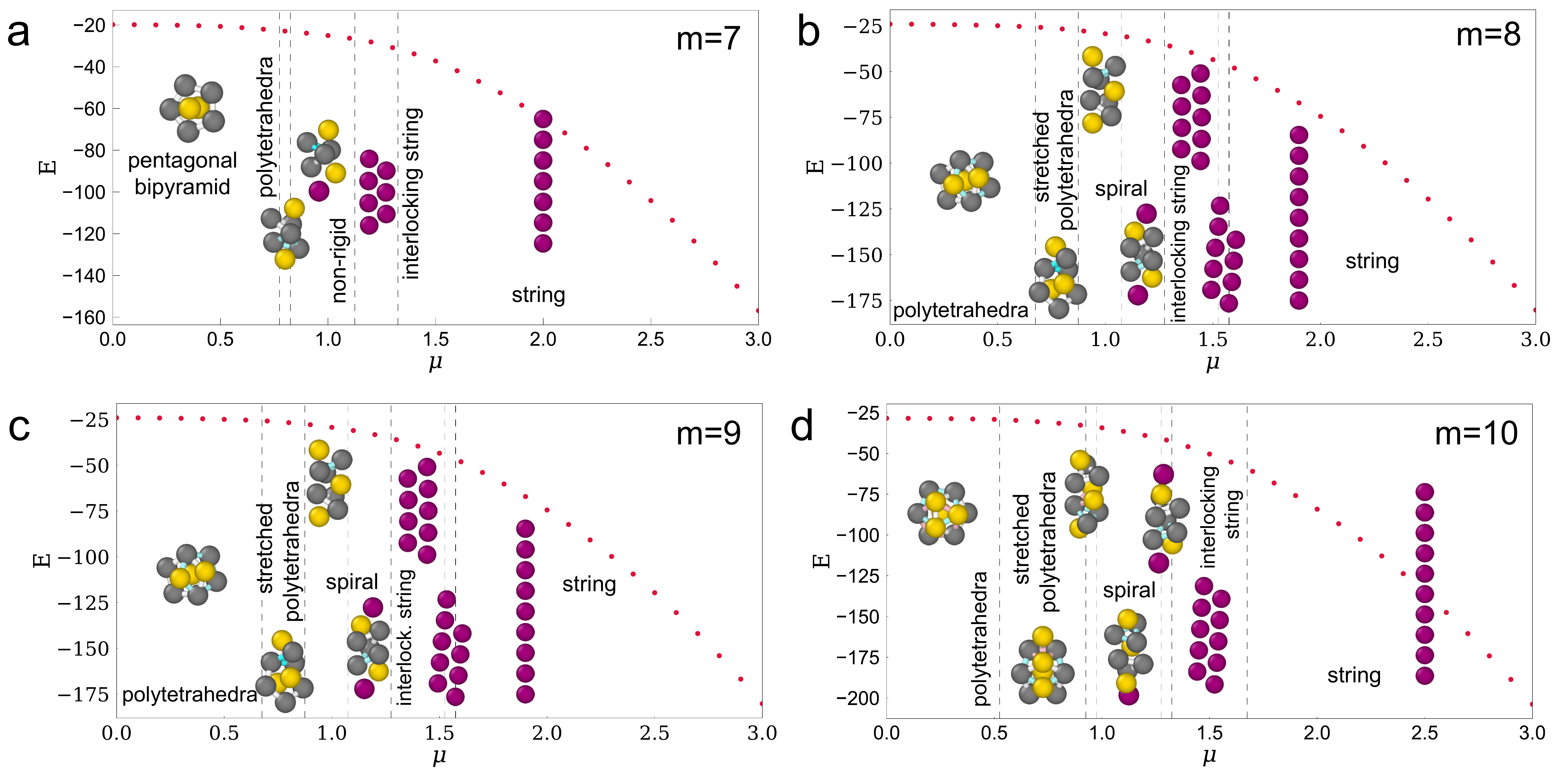}
\caption{Structural phase diagrams for $n=7$ to $m=10$.Representation of the minimum energy cluster of seven to ten dipolar particles as a function of dipole strength $\mu$, and their associated energy. Dashed lines indicate the change of structure. Dotted line is a best fit through the energy corresponding to each geometry.}
\label{figPhaseN7to10}
\end{figure*}

\begin{figure*}
\centering
\includegraphics[width=180 mm]{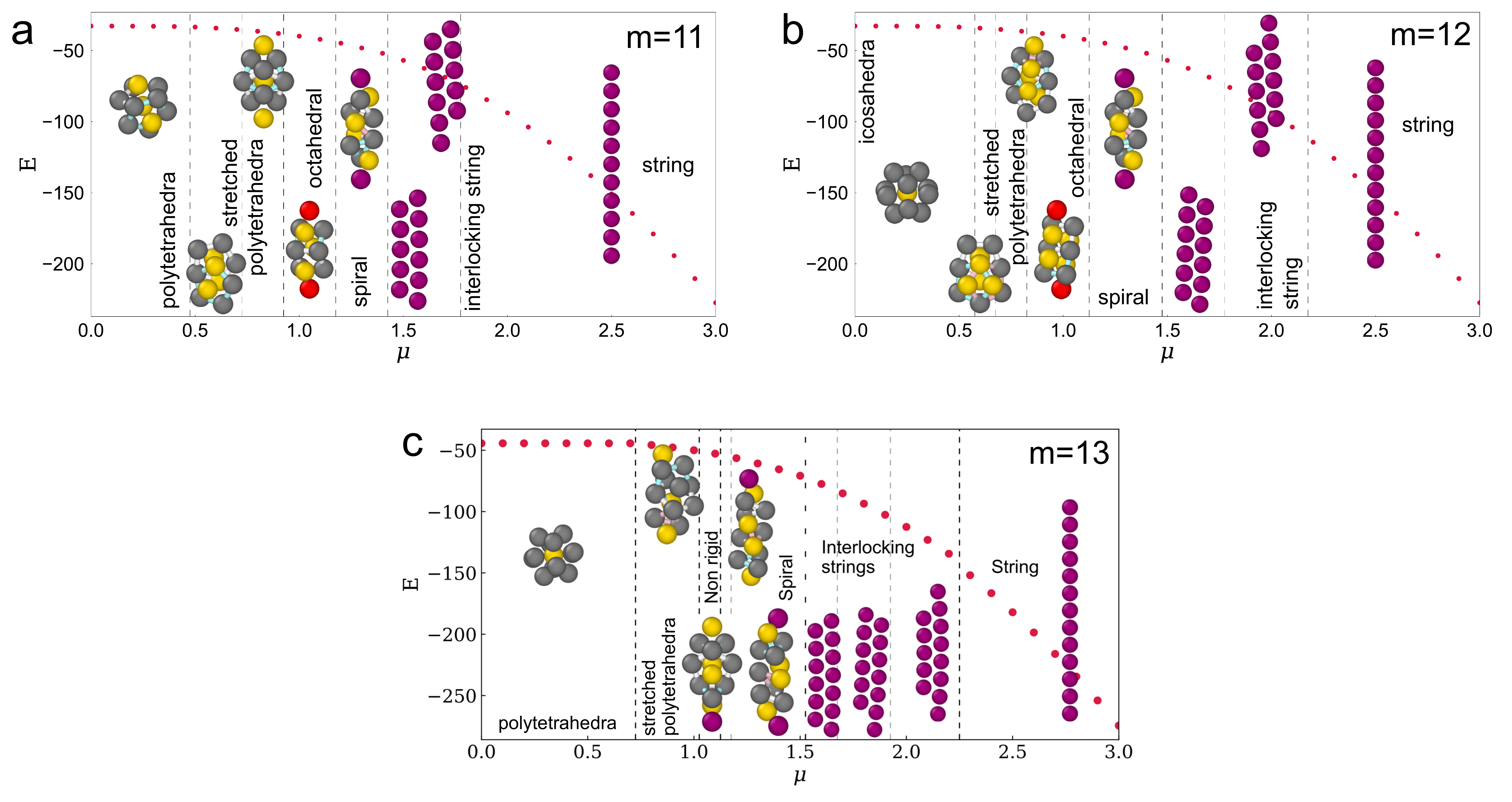}
\caption{Structural phase diagrams for $m=11$ to $m=13$. Representation of the minimum energy cluster of eleven to thirteen dipolar particles as a function of dipole strength $\mu$, and their associated energy. Dashed lines indicate the change of structure. Dotted line is a best fit through the energy corresponding to each geometry.}
\label{figPhaseN11to13}
\end{figure*}

\subsection{Molecular dynamics simulation}
\label{sectionMolecular}

We will show that in the Lennard--Jones--dipolar system a number of novel minimum energy clusters are found. Cluster populations are an important characteristic of colloidal fluids. In order to demonstrate the viability of these clusters as a characteristic of dipolar colloidal systems, we perform molecular dynamics simulations and use the topological cluster classification to identify the target clusters. Simulations were performed with the LAMMPS molecular dynamics package~\cite{plimpton1995} with Brownian dynamics time integration~\cite{moore2023}.

Simulations are performed on a system of 512 particles, at a effective volume fraction of around 0.0509 with the effective hard sphere diameter determined with the Barker-Henderson method where we consider the WCA part of the interaction potential to define the hard core~\cite{hansen}. Simulations are carried out for the following values of the dipole strength: \textmu=0, 0.3, 0.6, 0.9, 1.2, 1.5, 1.8, 2.1, 2.4, 2.7 (for each case the dipole strength is constant throughout the simulation). The simulations are run for $5\times10^8$ time units with a timestep of 0.00001. To investigate the phase behaviour we introduce an additional repulsive potential which tends to form clusters. In order to suppress aggregation (due to the Lennard--Jones interaction), and maintain the system as small clusters, we add a weak Yukawa potential to Eq.~\ref{eqULJdip} of the form:

\begin{equation}
\beta u_\mathrm{yuk}(r) =- \beta A\frac{e^{-\kappa r}}{r/\sigma}
\end{equation}

\noindent
where the prefactor $A=1$ and inverse screening length $\kappa \sigma=0.5$. The simulations are then evolved according to $\beta u_\mathrm{sim}(r,\theta) = \beta u_\mathrm{ljdip}(r,\theta) +\beta u_\mathrm{yuk}(r) $.

\begin{figure*}
\centering
\includegraphics[width=105mm]{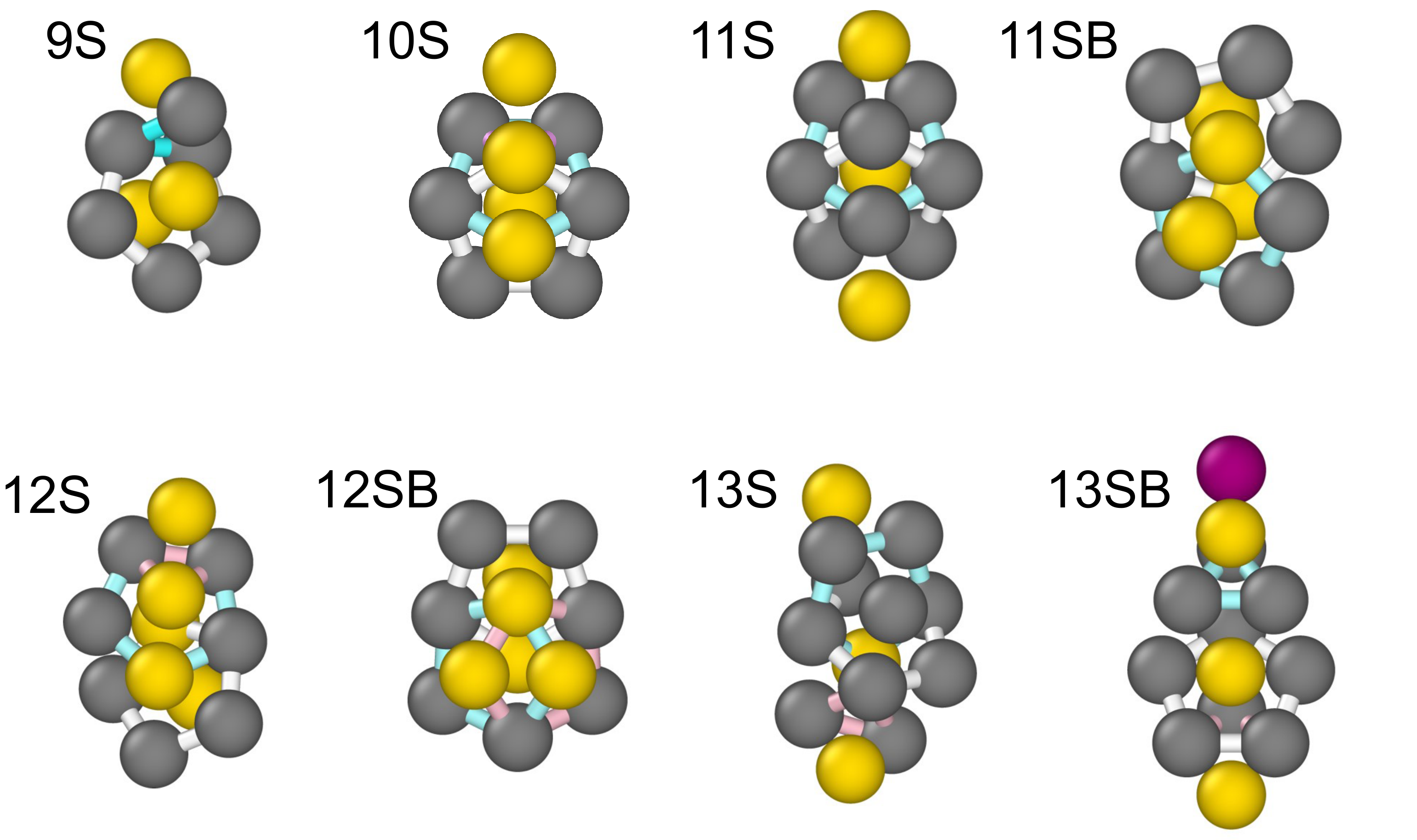}
\caption{Renderings of non--Lennard--Jones or ``stretched'' polytetrahedral clusters.
For the 9S, the five--membered ring is denoted with the white bonds and three--membered ring with blue bonds.
For the 10S, the interlocking five--membered rings are indicated with blue and white bonds. The three--membered ring towards the top of the cluster is indicated with pink bonds.
The two interlocking five--membered rings of the 11S and 11SB are shown in blue and white.
In the case of the 12S, the two five--membered rings are indicated in blue and white, the three membered ring is shown in pink.
For the 12SB, the three five--membered rings are indicated in white, blue and pink.
For the 13S, the interlocking five--membered rings are shown in blue and white.
The 13SB has a three--membered ring (blue) and a five--membered ring (white).
}
\label{figS}
\end{figure*}

\section{Results}
\label{sectionResults}

\subsection{Minimum energy clusters of the fixed dipolar--Lennard--Jones system}
\label{sectionMinimum}

\begin{figure}
\centering
\includegraphics[width=75mm]{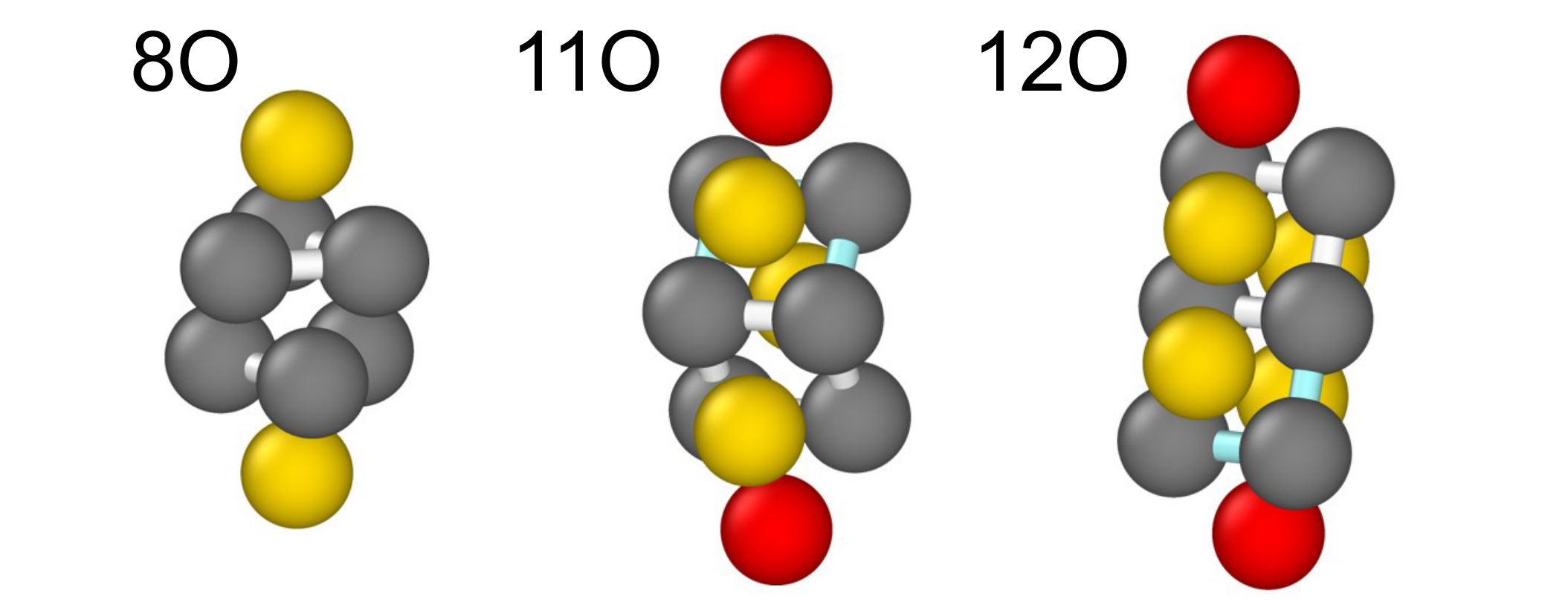}
\caption{Renderings of clusters based on the 6A octahedron.
The two three--membered rings are rendered with white bonds in the 8O cluster.
For the 11O, the two four--membered rings are indicated with white and blue bonds.
For the 12O, the four--membered rings are shown in blue and white.
}
\label{figO}
\end{figure}

The change in topology of minimum energy cluster as the cluster size and dipole strength are varied may be represented as a ``phase diagram'' in the energy $E$ -- dipole strength $\mu$ plane, as shown in Fig.~\ref{figGMINPhase}. For all cluster sizes, at high dipole strengths the minimum energy cluster is a string, and at low dipole strengths the minimum energy cluster we recover the polytetrahedral Lennard Jones minima. These have been previously determined as 7A, 8B, 9B, 10B, 11C, 12B and 13A ~\cite{wales1997} in the nomenclature of~\citet{doye1995} but are reproduced in Fig.~\ref{figLennardJones} in the Appendix for completeness. Before the strings form at high field strength, the particles form an interlocking string which before the regime of strings one particle wide is reached. Between these, the intermediate clusters may be grouped as polyetrahedral, clusters based on 6A octahedra (Fig.~\ref{figBasic}), and spiral clusters.

\begin{figure*}
\centering
\includegraphics[width=165mm]{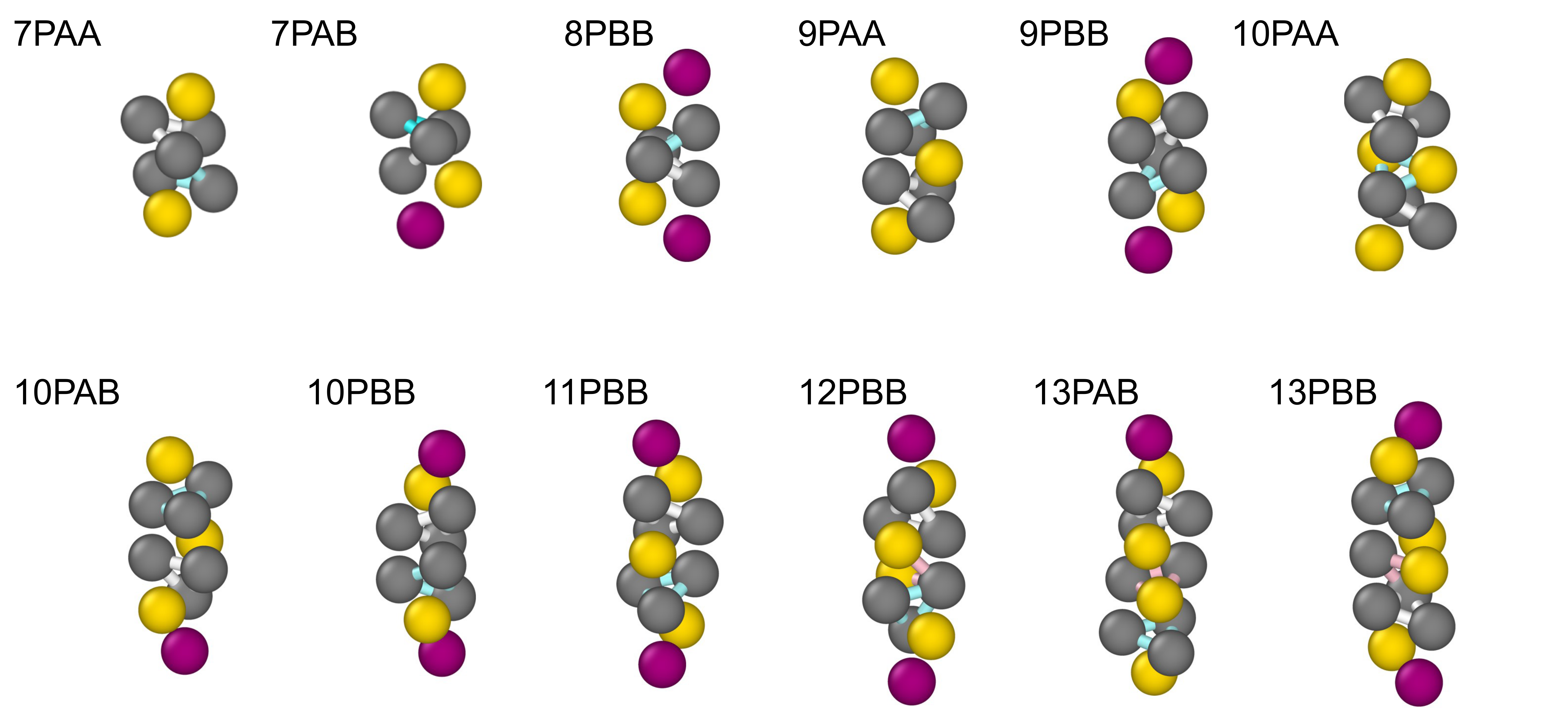}
\caption{Renderings of spiral clusters.
For the 7PAA,7PAB, 8PBB, 9PAA and 9PBB, the two three--membered rings are indicated with blue and white bonds.
In the case of the 10PAA, the three three--membered tins are indicated with blue and white bonds.
For the  10PAB and 10PBB, the two three--membered rings are indicated with blue and white bonds.
For the 11PBB, the two three--membered rings are indicated with white and blue bonds.
The 12PBB has three three membered rings shown with white, pink and blue bonds. 
For the  13PAB and 13PBB the three three membered rings are indicated with pink, white and blue bonds.
}
\label{figP}
\end{figure*}

\begin{figure}
\centering
\includegraphics[width=30mm]{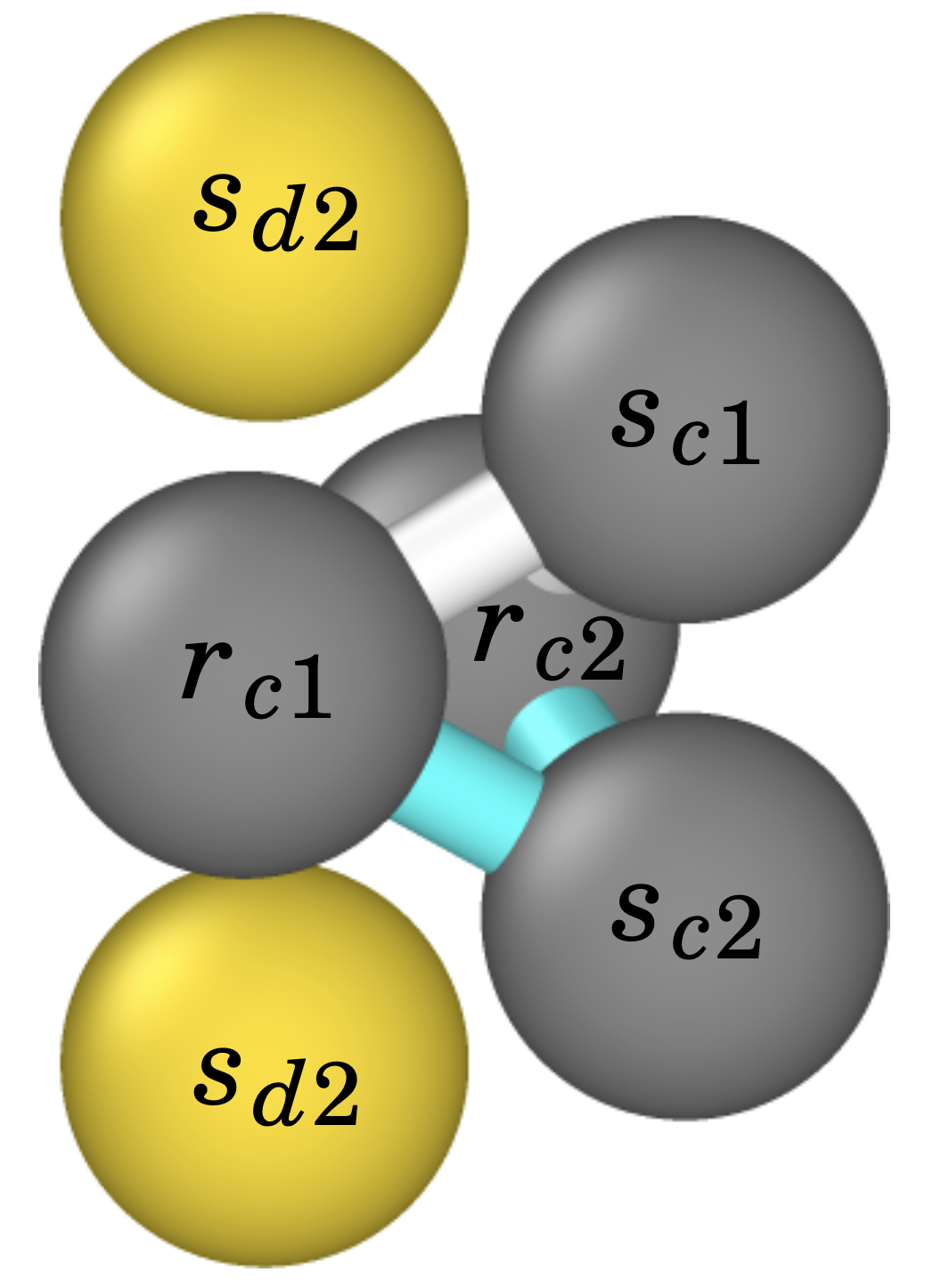}
\caption{6Z cluster with labelled particles. See text for the interpretation of the labels.}
\label{fig6Zlabelled}
\end{figure}

\begin{figure*}
\centering
\includegraphics[width= 125mm]{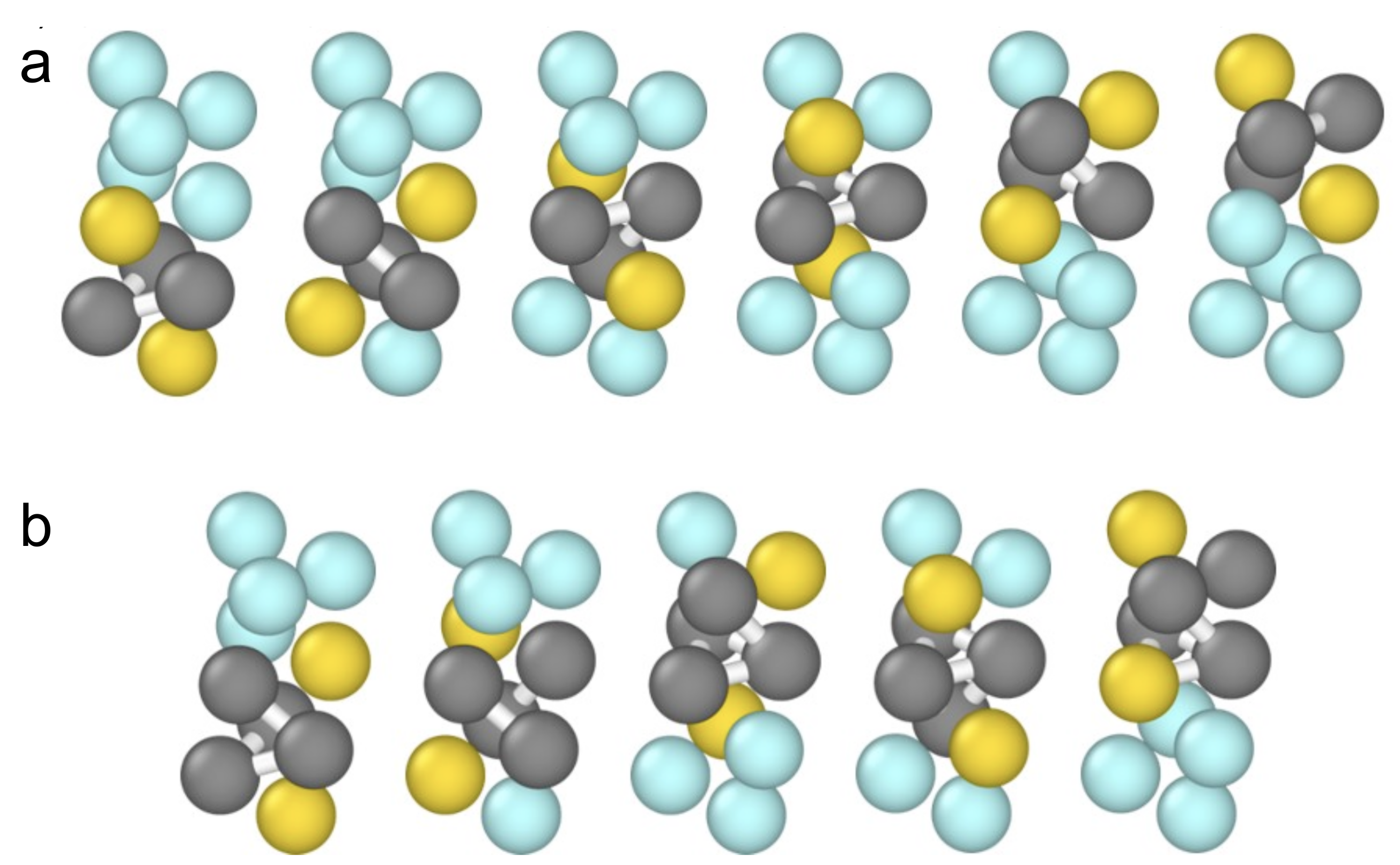}
\caption{Building the 10PAA cluster from 
(a) six intersecting 5A or equivalently
(b) five intersecting 6A clusters.
}
\label{figBuilding10PAA}
\end{figure*}

Figures~\ref{figPhaseN7to10} and ~\ref{figPhaseN11to13} show the full cluster ``phase diagrams'' for cluster sizes from 7 to 13 particles. As before in Fig.~\ref{figBasic}, the grey particles indicate rings and the yellow particles are spindles. Where a particle is both a ring and a spindle they are colored for clarity. Bonds indicate rings. Red particles are bonded to three particles as part of a single--spindle 4A cluster (Fig.~\ref{figBasic}). Purple particles are non--rigid.

\subsection{Identification of clusters with the Topological cluster classification}
\label{sectionIdentification}

Once the minimum energy clusters have been determined, we seek a means to identify them in bulk coordinate data, either in experiments or simulations. To this end, we have implemented the new clusters into the topological cluster classification. Table.~\ref{tabDetection} details the detection routines whereby each rigid cluster may be identified from simulation or experimental coordinate data. Figure ~\ref{figBasic} shows four structures which are the building blocks of all the larger rigid clusters. We now describe the detection procedure for each geometry -- stretched polytetrahedra, octahedra--based and spiral clusters.

\subsubsection{Non--Lennard Jones ``stretched'' polytetrahedra}
\label{sectionNonLennard}

As the dipole strength is increased from zero, the first class of clusters are non-LJ polytetrahedra. The dipolar interaction acts to elongate the cluster, moving one or more particles from the sides of the cluster to the top or bottom. We denote these clusters by ``S'' due to their stretching with respect to the $\mu=0$ case. For $m\ge11$, there is more than one polytetrahedral minimum before the system transitions to the spiral or octahedral cluster. These clusters are rendered in Fig.~\ref{figS}.

The 9S and 10S are formed by taking the Lennard-Jones cluster for 8 or 9 particles and adding an additional particle such that this particle forms a 5A triangular bipyramid which points in the direction of the field. The 11S and 12S clusters are formed from two and three intersecting 7A pentagonal bipyramids, which are stacked such that the cluster is elongated parallel to the field. The 11SB cluster is formed from the 9A cluster with two additional particles each forming a 5A at either end of the cluster. The 12SB cluster is formed from an 11S cluster with an additional particle added to form a 5A aligned with the dipoles. The 13S cluster is formed from a 12SB cluster with an additional particle forming a 5A. Finally, the 13SB is a 12S with an additional particle forming a three-membered ring. 13SB is the only non-rigid cluster that is not a spiral, an interlocked string or a string.

\subsubsection{Octahedra--based clusters}
\label{sectionOctahedral}

For $m=8$, $m=11$ and $m=12$, there is no point where the energy minimum is the non-rigid $m$PAB spiral (see Sec.~\ref{sectionSpiral}), instead, between the non-LJ polytetrahedra and the $m$PBB non-rigid spiral, the energy minimum cluster is based on the 6A octahedron. We therefore denote these clusters as ``O'': 8O, 11O, 12O (see Fig.~\ref{figO}). These clusters are intersections between 6A octahedra and 4A tetrahedra.tabDetection

8O is formed from a 6A cluster and two 4A clusters. 11O is formed by two intersecting 6A clusters which share two rings and one spindle particle. The remaining spindle and ring of each 6A form a 4A cluster with one additional particle. 12O is also formed from two 6A and two 4A clusters, however here the 6A clusters share two ring particles but no spindles. 11O and 12O therefore represent two types of octohedral clusters differentiated by the number of particles shared by the 6A. The next largest octahedral clusters of type 11O would be 14O, 17O, 21O, etc. The next largest clusters of type 12O would be 16O, 20O, 24O etc. The study of these larger octahedral clusters would be a useful target for future research.

\subsubsection{Spiral clusters}
\label{sectionSpiral}

For each cluster size studied, there is a range of values of dipolar strength $\mu$~where the minimum energy cluster is a spiral. For each value of $m$, there are three spirals denoted PAA, PAB and PBB, however only for $m=10$ are all three clusters an energy minima for some value of $\mu$. As shown in Fig.~\ref{figP}, the basinhopping simulations have identified the following clusters as energy minima for some value of $\mu$: 7PAA, 7PAB, 8PAB, 9PAA, 9PBB, 10PAA, 10PAB, 10PBB, 11PBB, 12PBB, 13PAB. The $m$PAA are fully rigid, $m$PAB have one non--rigid particle (purple in Fig.~\ref{figP}) and $m$PBB have two non--rigid particles. We note
that for N=8,11,12, the octahedral cluster would appear to replace the PAB cluster. Having a helical structure, the spiral clusters are chiral. The basinhopping outputted both right and left-handed spirals. These have the same energy and the TCC search algorithm does not differentiate between them.

We will first consider the rigid spirals 9PAA and 10PAA. The spiral $m$PAA cluster is comprised of $m-5$ intersecting 6Z clusters. 6Z is the minimum energy cluster of 6 particles interacting through the Dzugutov potential. Since 6Z is itself two intersecting 5A clusters, the spirals may equally be considered intersecting 5A. Figure~\ref{fig6Zlabelled} shows the labelled particles in a 6Z cluster. To be considered a 6Z, the two 5A clusters must intersect such that they share two ring particles $rc1$ and $rc2$. The remaining ring particle of each cluster must be a spindle particle of the other $sc1$ and $sc2$. The final two spindles are denoted $sd1$ and $sd2$.

Figure \ref{fig6Zlabelled} shows the constituent 5A and 6Z clusters for 10PAA. If one particle is removed from either end of the 10PAA, we have a 9PAA cluster comprised of four intersecting 6Z or five intersecting 5A. A 10PAA cluster adds another particle to this 9PAA such that the additional particle forms a spindle of another intersecting 5A cluster. This new 5A cluster must intersect with the lower or upper 5A cluster such that a new 6Z cluster is formed.

To build larger spirals, this same routine is performed at the ends of the spiral. In a 6Z cluster, there are only two 5A clusters, but for larger spirals, there will be more, only the 5A clusters at the top and bottom of the cluster can grow the cluster by forming a new 6Z. Therefore in the implementation of the topological cluster classification, each particle must be assigned its own index.

The other class of spirals are the non--rigid spirals. In these clusters, an additional particle is added to a spiral which is bonded in a three--membered ring to the non-shared spindle and the non--shared ring particle of the outermost 6A. In the implementation of the topological cluster classification, this third particle must have only two bonds to the existing cluster.

\begin{figure}
\includegraphics[width=\linewidth]{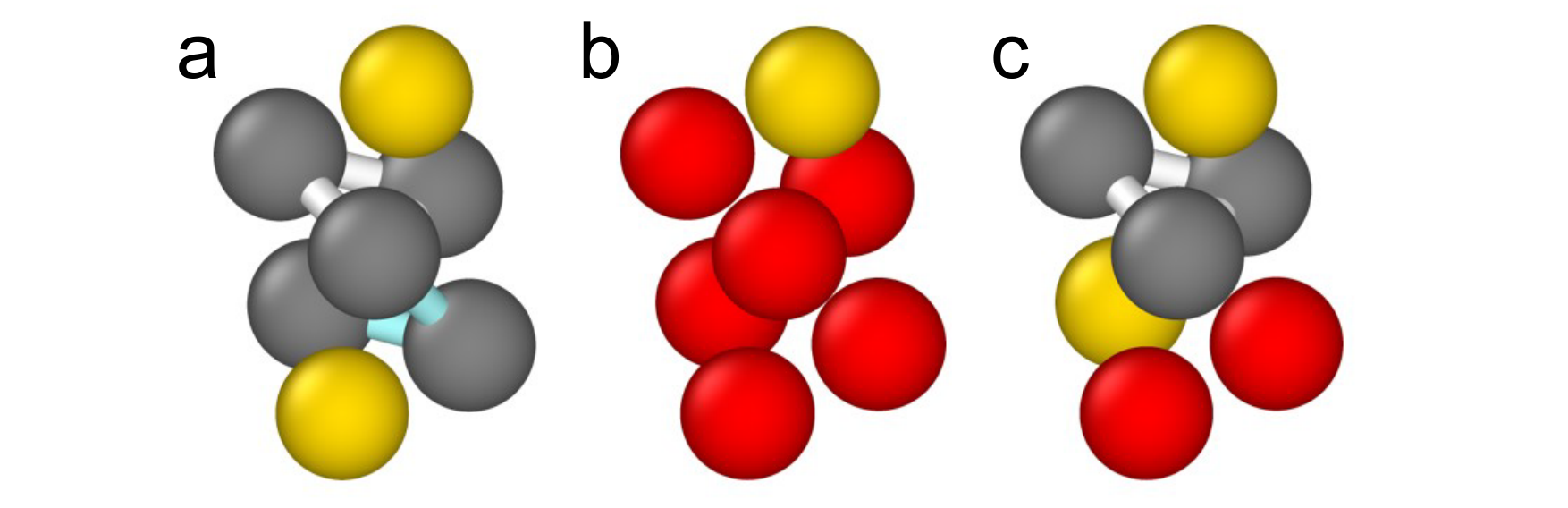}
\caption{Adding particles to grow a spiral 7PAA cluster from a 6Z. 
(a) shows a 7PAA as two intersecting 5A clusters. b) shows the 6Z cluster and the additional particle. c) shows how the additional particle forms a 5A with the original 6Z cluster.}
\label{fig7A6Z}
\end{figure}

\subsection{Results from molecular dynamics simulations}
\label{sectionResultsMolecular}

\begin{figure}
\centering
\includegraphics[width=\linewidth]{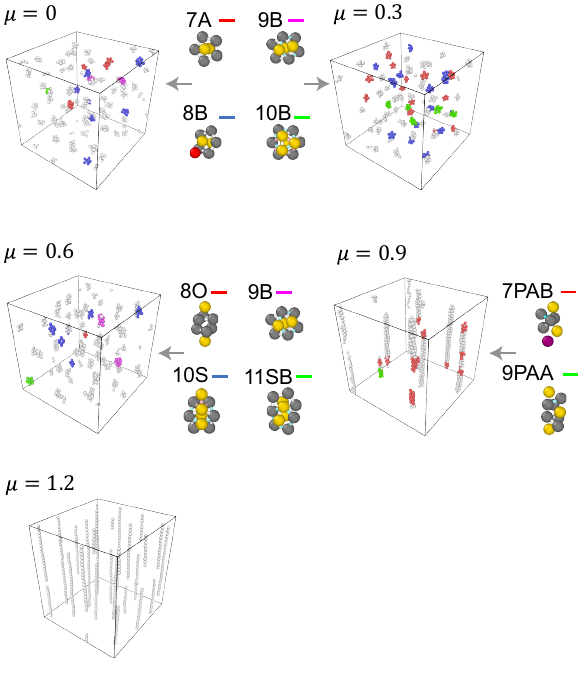}
\caption{Rendering of molecular dynamics simulation, with corresponding minimum energy clusters shown. These snapshots were taken at the end of the run, ie after $10^4$ time units.
The colours of the particles correspond to the lines by the clusters depicted to the right of the snapshots.}
\label{figYukawaRendering}
\end{figure}

\begin{figure*}
\centering
\includegraphics[width=.8\linewidth]{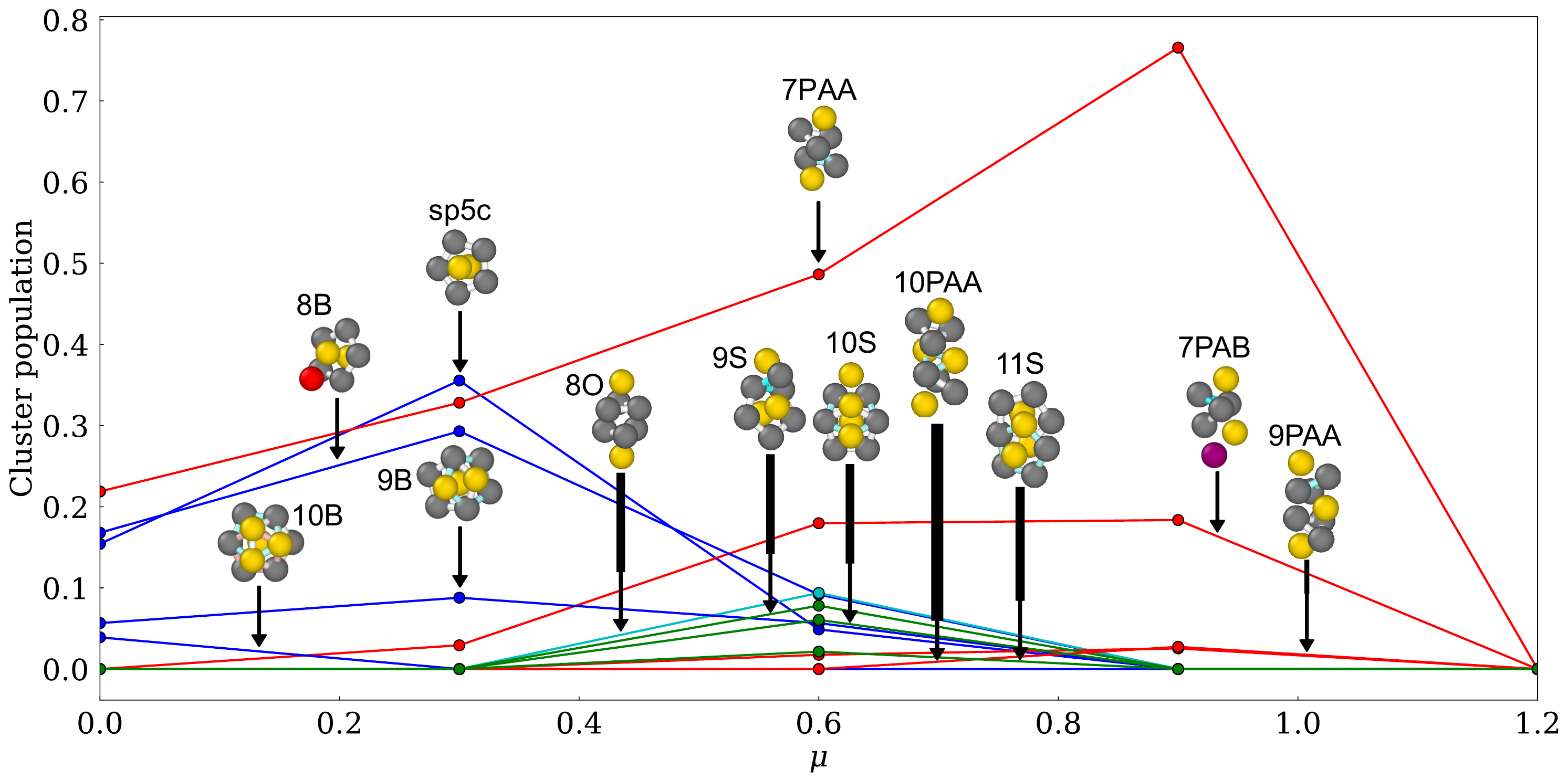}
\caption{Population of clusters as a function of dipole strength in molecular dynamics simulation data. }
\label{figYukawa}
\end{figure*}

The populations of all the minimum energy clusters obtained from the molecular dynamics simulations performed as described in Sec.~\ref{sectionMolecular} will now be considered for various values of the dipole strength $\mu$. Snapshots are shown in Fig.~\ref{figYukawaRendering} for certain values of $\mu$.

The results, shown in Fig. \ref{figYukawa} show the populations of all the clusters, grouped by size, as a function of $\mu$. This shows that the populations of the different classes of clusters show qualitative agreement with the transitions shown in the previous section. At low dipole strengths, the populations of Lennard-Jones clusters are highest. The populations of the non-LJ stretched polytetrahedra are also higher at the low dipole strengths, and in the cases of $m=10,11,12$ and $13$, the non-LJ polytetrahedra persist to higher field strengths than the Lennard-Jones clusters. The results show that the peak in the population of the 8O and 12O octahedra--based clusters is found at intermediate field strengths. These clusters are both observed within larger octahedral columnar structures. The absence of the 11O cluster is suggested as an avenue for future research. Finally, the populations of smaller spiral clusters 7PAA, 7PAB, and 9PBB peak at $\mu \approx 1$. This is at a higher value of $\mu$~than the octahedral clusters, which is qualitatively consistent with the transition found in the basin--hopping simulations. It may be expected that for larger clusters, the transitions shift to higher values of $\mu$. Therefore, whilst the peaks in octahedral, non-LJ polytetrahedra and spiral clusters are at higher values of $\mu$~than in the GMIN simulation results, these still support a transition from Lennard-Jones to non-LJ polytetrahedra to octahedra--based to spirals as the dipole strength is increased. The results also imply that the order of these transitions would hold for larger clusters.

\section{Discussion}
\label{sectionDiscussion}

Helical structures at the micro and nanoscale have been suggested as promising structures in opto-electronics~\cite{rosen2011} and catalysis~\cite{Gier1998}, and are for understanding chiral structure in biopolymers~\cite{pijper2008,yashima2009}. The assembly of Bernal spirals is also seen in systems of patchy colloids~\cite{morgan2013,fejer2014}
In addition, string structures comprised of stacking tetrahedra have also been observed in polarised, metallodielectric Janus particles~\cite{gangwal2005,hong2008} and notably the active colloidal dipolar system~\cite{sakai2020}. The observation of helical structures in the potential energy landscape is therefore of importance to a wide range of systems. It is quite possible that many of the same structures we have identified here would be found in these systems, which would be interesting to explore in the future.

In addition to these systems with anisotropic interactions, some of the clusters that we have identified here have in fact been observed also in systems with \emph{isotropic} interactions. In particular, so--called short--range attraction long--range repulsion or SALR particles~\cite{royall2018thunderbox} have been shown to form Bernal spirals~\cite{campbell2005}. Indeed these are consistent with minimum energy clusters determined for SALR potential~\cite{mossa2004}. In the future it would be an interesting extension of this work to investigate the effect of weak repulsions in perturbing the clusters from isotropic Lennard--Jones minima, along with the other systems anisotropic interactions that we have mentioned here.

While the computer simulations we have carried out merely serve as a proof--of--principle of the method we have implemented, we nevertheless take the liberty to make a few comments. As noted, the overall trends are very much in agreement with the expectation of Lennard--Jones polytetrahedra, stretched polytetrahedra, octahedra-based, spiral clusters and strings as a function of dipolar strength. However in comparison to work which considered a system with spherically symmetric interaction~\cite{malins2010}, there appear to be a rather larger number of particles not identified in a minimum energy cluster. Whether this is due to geometric frustration~\cite{malins2009} or some other cause would be an interesting avenue for future research.

\section{Conclusion}
\label{sectionConclusion}

We have identified a library of minimum energy clusters of a particles with a Lennard--Jones interaction with a dipole fixed in one direction. We consider cluster sizes $7 \le m \le 13$. This model is relevant to colloidal systems in which dipolar interactions are induced by an electric field. These turn out to to exhibit a rich structural ``phase diagram'' with a variety of topologies. As the dipolar strength is increased, the clusters transition from (relatively) isotropic polytetrahedra which are minimum energy clusters of the Lennard--Jones model. These transition to ``stretched'' polytetrahedra, clusters based on octahedra, to Bernal spirals to interlocking before finally forming strings at high dipole strength.

We have implemented a search for these clusters in the topological cluster classification~\cite{malins2013tcc}. Therefore it is now possible to analyze systems with anisotropically interacting particles in the same way as isotropically interacting systems were previously analyzed by carrying out molecular dynamics simulations. Finally, we provide an example of such analysis with a molecular dynamics simulation for a system whose interactions are very similar to the Lennard--Jones--dipolar interaction. We see a progression from isotropic Lennard--Jones clusters through the sequence of geometries to strings.

\begin{table*}
\begin{ruledtabular}
\begin{tabular}{llc}
Cluster & Detection routine & Figure \\ \hline
8O      & A 6Z cluster and two 4A clusters where: & ~\ref{figO} \\ 
           & Each 4A cluster shares three particles with the 6Z and the two 4A clusters have no particles in common & \\ \hline
9S      & An 8B and a 5A where: &  ~\ref{figS} \\ 
          & The clusters share a spindle particle and two ring particles.  & \\ 
          & The remaining ring particle of the 5A is the ``additional'' (not part of a 7A) particle of the 8B. & \\ \hline
9PAA & Two 6Z clusters and one 5A cluster where: & ~\ref{figP} \\
          & The 6Z clusters share a common spindle and two common ring particles. & \\
           & The remaining spindle of each 6Z cluster is a ring particle of the other. & \\
           & The 5A shares a common spindle particle with  6Z$_i$. & \\
           & The 5A shares two ring particles with 6Z$_i$.  The remaining 5A ring particle is the remaining spindle 6Z$_i$. & \\ 
           & All the particles in  5A and 6Z$_j$ are distinct  \\ \hline
10S      & A 7A and a 5A cluster where: & ~\ref{figS} \\
            & The clusters have one common spindle particle and one common ring particle. &  \\
            & The clusters have only two particles in common. &  \\ \hline
10PAA & Two 9PAA clusters which have 8 particles in common & ~\ref{figP} \\ \hline
11S      & Three 7A clusters where:  & ~\ref{figS} \\
            & 7A$_i$ and 7A$_j$ form a 9A. & \\
            & The ring of 7A$_k$ is the uncommon spindle of 7A$_j$ and an uncommon ring of 7A$_j$. & \\
            & The ring particles of 7A$_k$ comprise: the other uncommon ring of 7A$_j$, & \\ 
            & the common spindle particles of 7A$_i$ and 7A$_j$, one common ring particle of 7A$_i$ and 7A$_j$& \\  \hline
  11SB & two 7A and two 5A clusters where:  & ~\ref{figS} \\
            & 7A$_i$ and 7A$_j$ form a 9A and 5A$_i$ and 5A$_j$ have a common spindle. & \\
            & One ring particle of 5A$_{ij}$ is a spindle of 7A$_{ij}$, and this spindle is not the 7A$_{ij}$ shared spindle. & \\
            & Two ring particles of 5A$_{ij}$ are ring particles of 7A$_{ij}$, and these ring particles are not shared between 7A$_{ij}$. & \\ \hline  
  11O   & Two 6A clusters and two sp3b clusters where: & ~\ref{figO} \\
            & 6A$_i$ and 6A$_j$ have two particles in common. & \\
            & sp3b$_{ij}$ has three common particles with 6A$_{ij}$ and none with 6A$_{ji}$. & \\
            & There is no sp3b$_k$ within 6A$_i$ and 6A$_j$ which has common particles with both sp3b$_i$ and sp3b$_j$. & \\ \hline
  12S   & Three 7A clusters where: & ~\ref{figS} \\
            & 7A$_i$ and 7A$_j$ share 3 particles. 7A$_i$ and 7A$_k$ share 3 particles. & \\
            & 7A$_k$ and 7A$_j$ share 4 particles. 7A$_i$ has a spindle particle which is a ring of both 7A$_j$ and 7A$_k$. & \\
            & 7A$_i$ has a ring particle which is a spindle of both 7A$_j$ and 7A$_k$. Any two 7A have a common ring particle. & \\ \hline      
  12SB & A 11SB and a 5A where: & ~\ref{figS} \\
            & One spindle of the 5A is the common spindle particles of the two 7A clusters within the 11SB.& \\
            & Three particles of the 5A are common with one of the 7A clusters within the 11SB. & \\ \hline
  12O   & Two 6Z and two sp3b clusters where: & ~\ref{figO} \\
            & 6Z$_i$ and 6Z$_j$ share two particles. sp3b$_{ij}$ has three common particles with 6Z$_{ij}$ and none with 6Z$_{ji}$. & \\
            & There is no sp3b$_k$ within 6A$_i$ and 6A$_j$ which has common particles with both sp3b$_i$ and sp3b$_j$. & \\ \hline
  13S    & A 12SB and a 5A cluster where: &  ~\ref{figS} \\
            & One ring of the 5A is the "additional" (not part of an 11SB) particle of the 12SB. & \\
            & One ring of the 5A is an "additional" (not part of a 9B) particle of the 11SB within the 12SB.& \\ 
            & The remaining ring particle of the 5A is common with the 12SB. & \\ \hline
\end{tabular}
\end{ruledtabular}
\caption{Detection routines for all \emph{rigid} minimum energy clusters of of Dipolar particles whose dipole is fixed with respect to an external axis. Here ``P'' clusters are spiral, ``S'' clusters are polytetrahedral but ``stretched'' with respect to the Lennard--Jones clusters and ``O'' clusters are based on 6A octahedra. The letters "AA" in the spiral cluster name indicate that both ends of the cluster are rigid "AB" or "BB" would indicate that one or both ends was non rigid. Detection for the spiral clusters $m$PAA, $m$PAB and $m$PBB is given in Sec.~\ref{sectionSpiral}.}
\label{tabDetection}
\end{table*}

\section*{Appendix}
\setcounter{figure}{0}
\renewcommand{\thefigure}{A\arabic{figure}}

\begin{figure*}
\includegraphics[width=120mm]{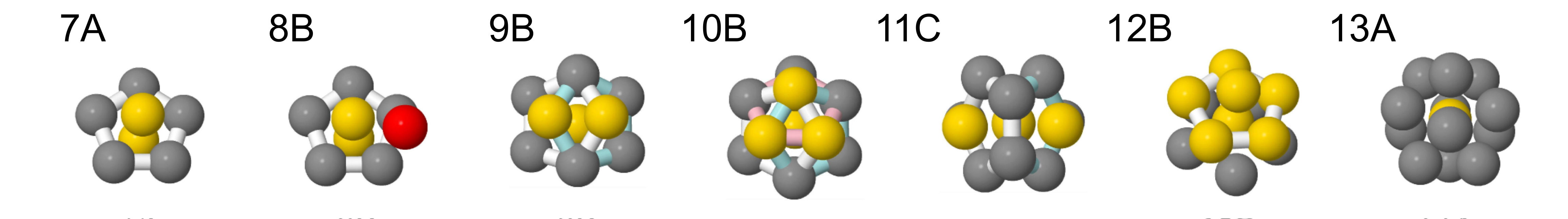}
\caption{Lennard--Jones minimum energy clusters, determined by~\citet{wales1997}. These have been implemented in the TCC previously~\cite{malins2013tcc}.}
\label{figLennardJones}
\end{figure*}

\begin{acknowledgments}
KS thanks the EPSRC doctroal training program for support.
CPR acknowledges the Agence Nationale de Recherche for grant DiViNew.
FJ.M was supported by a studentship provided by the Bristol Centre for Functional Nanomaterials (EPSRC Grant No. EP/L016648/1). 
The authors would like to thank Mark Miller, Josh Robinson and Peter Crowther for the help in building the TCC software. 
\end{acknowledgments}

\end{document}